\documentstyle[11pt,newpasp]{article}
\newcommand{\kms}{~{\rm km\,s^{-1}}}

\def\beq{\begin{equation}}
\def\eeq{\end{equation}}
\def\ben{\begin{eqnarray}}
\def\een{\end{eqnarray}}

\def\spose#1{\hbox to 0pt{#1\hss}}
\def\lta{\mathrel{\spose{\lower 3pt\hbox{$\mathchar"218$}}
     \raise 2.0pt\hbox{$\mathchar"13C$}}}
\def\gta{\mathrel{\spose{\lower 3pt\hbox{$\mathchar"218$}}
     \raise 2.0pt\hbox{$\mathchar"13E$}}}

\def\kms{{\rm\,km\,s^{-1}}}
\def\kpc{{\rm\,kpc}}
\def\Mpc{{\rm\,Mpc}}
\def\msun{{\rm\,M_\odot}}
\def\lsun{{\rm\,L_\odot}}

\def\yr{{\rm\,yr}}
\def\Gyr{{\rm\,Gyr}}

\def\d{{\rm d}}

\def\aa#1 #2 {AA, #1, #2}
\def\aasupp#1 #2 {AASuppl, #1, #2}
\def\apj#1 #2 {ApJ, #1, #2}
\def\apjsupp#1 #2 {ApJSupp, #1, #2}
\def\aj#1 #2 {AJ, #1, #2}
\def\mn#1 #2 {MN, #1, #2}
\begin{document}
\title{Accretion by Galaxies}
\author{J.J.~Binney}
\affil{Theoretical Physics, Keble Road, Oxford OX1 3NP, UK}

\begin{abstract}
Both theory and  observation indicate that galaxies like the Milky Way
accrete matter at the rate of a few M$_\odot$ per year.
\end{abstract}

\keywords{Galactic Structure, cosmology}

\section{Introduction}

The Milky Way is an extremely typical galaxy in the sense that most of the
luminosity in the Universe comes from similar systems. Therefore,
indications that the Milky Way is accreting at a significant rate would
imply that accretion is an important phenomenon within the Universe at
large.

Several observations suggest that material is falling into the Milky Way.
First and foremost, our nearest large neighbour, M31, is approaching us at
$\sim140\kms$. Second, the so-called `high-velocity clouds' have, on the
average, negative line-of-sight velocities. Third, the Magellanic Clouds are
on an orbit that must be decaying through dynamical friction, and I shall
argue that the orbit of the Sagittarius Dwarf Galaxy is in a still more
advanced state of decay. Fourth, beyond $\sim R_0$ the Galactic disk has a
complex warped structure, and the most plausible explanation of this
phenomenon invokes infall as the driver. Fifth, at a given radius the ISM
appears to vary significantly in metallicity from place to place, and this
phenomenon is most naturally explained by persistent accretion of
low-metallicity material.

In this talk I shall review each of these lines of evidence in turn.

\section{Collapse of the Local Group}

The luminosity of the Local Group is dominated by the Milky Way and M31.
Since M31 is approaching us, the Local Group is either virialized or
collapsing. The simplest dynamical model in which the Group's mass is
concentrated in point particles coincident with the Milky Way and M31
suffices to show that the age of the Universe, combined with the present
distance and Galactocentric velocity of M31, implies that the Local Group is
collapsing for the first time (Kahn \& Woltjer 1959). 

When a cosmological structure collapses for the first time, its constituent
parts move on highly elongated orbits towards the barycentre. Some of these
orbits must enter the Galactic halo and become trapped. How much matter is
the Galaxy likely to be accreting in this way?

The answer depends on two related quantities: how far out the Galactic halo
reaches, and what fraction of the Local group lies outside M31 and the
Galaxy. The mass of the Local Group has been estimated many times and found
to lie in the range\footnote{The Local Group's $V$-band luminosity is
$\sim4.2\times10^{10}\lsun$ (e.g., Tables 2.1 and 4.3 of Binney \&
Merrifield, 1998), while the mean mass-to-light ratio of the Universe is
$400h(\Omega/0.25)$ (\S10.3.1 of Binney \& Tremaine, 1987), so for $h=0.65$
one expects the mass of the Local Group to be $1.1\times10^{13}$.} $4$ to
$8\times10^{12}\msun$ (e.g., Schmoldt \& Saha 1998). If {\it all\/} this
matter were in M31 and the Galaxy, with a third of the matter being in the
Galaxy, the halo would have to extend at constant $v_c=220\kms$ to $120$ to
$210\kpc$.  Consequently, material that comes within $\sim100\kpc$ of the
Galactic centre is likely to be accreted. 

Blitz et al.~(1999) describe a simulation of the formation of the Local
Group in which $10^6$ test particles move in the field of two point masses
representing M31 and the Galaxy, plus the tidal field generated by external
galaxies from Raychaudhury \& Lynden-Bell (1989). Particles that come within
$100$ comoving kpc of either M31 or the Galaxy are captured. In the
simulation the Hubble flow currently reverses $\sim1.5\Mpc$ from the centre
of the Local Group and the Galaxy's current accretion rate is
$\sim7.5\msun\yr^{-1}$, of which $\sim0.8\msun\yr^{-1}$ takes the form of
neutral hydrogen. 

Can we see infalling material?  One school of thought has long held that the
so-called high-velocity clouds (HVCs) are made of infalling gas (Oort 1966, 1970). HVCs were
first identified when Muller, Oort \& Raimond (1963) detected 21-cm emission
in many directions at velocities that are incompatible with circular
rotation.  The location of the objects responsible for this emission has been
controversial, however. One possibility is that they are small, nearby
clouds that have been accelerated to large peculiar velocities by
supernovae, stellar winds, and the like. Alternatively, they might be at
distances in excess of a Mpc and be tracing the (disturbed) Hubble flow
around the Local Group. 

It is likely that no single explanation applies to all HVCs. Some of these
objects are almost certainly associated with Local-Group galaxies such as
M31, the Magellanic Clouds and the Phoenix dwarf spheroidal (St-Germain et
al, 1999). Others, such as Complex A (van Woerden et al., 1998) and Complex
M (Danly et al., 1993) are small clouds in the Galactic halo. But a powerful
case can be mounted that many HVCs are systems over $10\kpc$ in diameter
that lie at distances $\sim1\Mpc$.

Braun \& Burton (1999) identified a sample of 66 compact, isolated HVCs, 23
of which had not previously been catalogued. They showed that these objects
are distributed fairly uniformly over the sky, define a mean velocity that,
within the errors, agrees with the Local Group's mean velocity (Karachentsev
\& Makarov, 1966), and with respect to this mean have a velocity dispersion
of $69\kms$ and an infall velocity $\sim100\kms$.  They infer that these
objects are typically $\sim15\kpc$ in diameter and contain $\gta10^7\msun$
of HI.

Blitz et al.~(1999) point out first that HVCs are seen within the Galactic
plane where local objects would collide at high velocity with gas in
circular motion, yet there is no evidence for shock-excited gas. Second,
attempts to detect HVCs in absorption against distant stars have been
largely fruitless -- only two HVCs have been detected in absorption, and
there are other reasons for believing that these are two of three HVCs that
are unusually nearby. Third, nearby clouds would be illuminated by UV
photons from the Galaxy and be brighter H$\alpha$ sources than they actually
are. Fourth, the velocities of the HVCs make more sense when referred to the
barycentre of the Local Group than when referred to either the velocity of
the Galactic centre or the Local Standard of Rest. Fifth, the metallicities
of HVCs are low, which is inconsistent with their having been ejected from
the Galactic ISM.

Blitz et al.\ use their simulation of the formation of the Local Group to
show that the non-uniform distribution of HVCs on the sky arises naturally
if HVCs lie at distances $\gta1\Mpc$. Moreover, at such distances the HVCs
would have a distribution over column density which resembles that of
Ly$\alpha$ clouds, which are now known to be floating in intergalactic
space (Theuns \& Efstathiou, 1998; Dav\'e et al.\ 1999). Hence, when the HVCs are interpreted as extragalactic objects, they
fit naturally into the picture of intergalactic space that has emerged from
a combination of theoretical and observational work in cosmology. Blitz et
al.\ argue that objects lying near the top of the HVC mass spectrum have
already been detected in external groups of galaxies, and that less massive
objects are  detectable with feasibly deep observations. They estimate that
a typical cloud contains $3\times10^7\msun$ of HI, so the population as a
whole contains $\sim10^{10}\msun$ of HI.

Once one accepts that many HVCs are extragalactic and associated with cosmic
infall, it is inevitable that their detected hydrogen is associated with a
very much larger mass of dark matter.  The estimate of the Galaxy's
accretion rate cited above is based on the assumption that there is 10 times
as much dark matter as neutral hydrogen. There could easily be more dark
matter by a factor of a few, since $\Omega_{\rm baryon}$ is thought to be
$\lta0.05$ and only a fraction of a HVC's hydrogen content will be neutral.

\section{Warps}

Both theoretical and observational developments over the last few years have
strengthened the argument that warps such as that of the Galactic disk are a
reflection of cosmic infall, and the associated reorientation of galactic
angular-momentum vectors. Since the discovery of the Galactic warp by Burke
(1957) and Kerr (1957), there has been a debate as to whether warps are
self-consistent intrinsic structures, or externally driven. Kahn \& Woltjer
(1959) suggested that the Galactic warp was driven by an intergalactic wind,
while Lynden-Bell (1965) suggested that a warp can be an intrinsic
structure.  The latter proposal was shot down by Hunter \& Toomre (1969).
After the discovery of dark matter, Toomre (1983) and Dekel \& Shlosman (1983) argued
that warps might after all be intrinsic structures, and Sparke \& Casertano
(1988) developed this proposal to the point that it became very attractive.
Recently Nelson \& Tremaine (1995) and Binney, Jiang \& Dutta (1998) have
demonstrated that the Toomre--Dekel proposal, as elaborated by Sparke \&
Casertano, is not viable. The flaw in the approach of Sparke \& Casertano is
the treatment of the dark halo as a rigid, unflexing thing. The
semi-analytic work of Nelson \& Tremaine and the numerical simulations of
Binney et al.\ demonstrate that a dark halo responds rapidly to any warp in
an embedded disk in such a manner that warps predicted to endure for ever by
Sparke \& Casertano actually wind up within a few dynamic times.

Binney et al.\ did not demonstrate that intrinsic warps are impossible, but
only that any long-lived warp would have to be a manifestation of a
cooperative distortion of {\it both\/} disk and halo. However, a priori it
is not clear that such configurations exist, and extensive numerical
experimentation by several groups has failed to find one. Hence, the
theoretical case for intrinsic warps must be considered at best doubtful.
Moreover, new observations (Shang et al.\ 1998) of the classic example of an
isolated galaxy with an integral-sign warp, NGC 5907, have shattered the
observational case for the existence of warps in isolated galaxies: NGC 5907
possesses both a dwarf satellite $\sim37\kpc$ from its centre, and an
elliptical ring of luminosity that probably contains the debris of another,
now shredded, companion.

What is the theoretical status of externally driven warps? The original
proposal of Kahn \& Woltjer (1959) is not viable because it relies upon a
massive, subsonic wind past the Milky Way, for which there is no evidence.
Similarly, proposals involving magnetic fields (Battaner, Florido \&
S\'anchez-Saavedra, 1990) are for various
reasons not in serious contention. An idea that remains plausible is that
warps are a response to the accretion by a galaxy of material laden with
angular momentum about an axis that is inclined to the galaxy's original
spin axis (Binney \& May, 1986; Ostriker \& Binney, 1989). Jiang \& Binney
(1999) followed the dynamics of a disk embedded in a live halo by
decomposing the disk into a series of massive rings that interact
gravitationally with one another and with the $100\,000$ particles that
represent the halo. The disk is exponential out to $R=3.5R_\d$ and is then
smoothly tapered to zero surface density at $R=4R_\d$. The halo initially has
ten times the mass of the disk and ensures that the overall rotation curve
is very nearly flat out to $5R_\d$. The accretion of material by the halo is
simulated by injecting particles into a torus of major radius $8.9R_\d$
whose spin axis is inclined by $15\deg$ to the original spin axis of the
disk. In response to the accretion, the disk develops an integral-sign warp
that carries the outermost ring $\sim0.3R_\d$ above and below the plane of
the innermost ring. Hence the warp is very comparable in magnitude to the
warp of the Milky Way. This simulation shows very clearly that the warp is
essentially a halo phenomenon: accretion causes the outer halo to tip with
respect to the inner halo. The disk largely acts as a tracer of the internal
dynamics of the halo. 

How much infalling matter is needed to generate a warp?
In this simulation, the reorientation of the galactic angular momentum is
driven by the crude expedient of adding mass to a tilted annulus of fixed
radius.  This procedure is well defined and easy to describe, but it
exaggerates the mass of infalling material that is required to slew a
galaxy's angular momentum by a given amount because neither the
angular-momentum per unit mass of the infalling material, nor the angle
between its spin axis and the galaxy's original spin axis increases with
time, as they would in a more realistic situation. Indeed, if the simulation
were continued for longer, the galaxy's angular momentum would
everywhere become parallel to the symmetry axis of the torus, and the warp
would fade away. In the real world, by contrast, the angular momentum vector
of infalling material will be constantly shifting its direction, and,
moreover, less and less material will be required to import a given quantity
of angular momentum. Hence the simulation's plausibility hinges on whether
the rate of angular-momentum slewing in it is plausible, rather
than on the likelihood that a galaxy will accrete as much material at
$\sim10R_\d$ as is crudely assumed.

Several authors have studied how tidal interactions endow protogalactic
regions with angular momentum at early times (e.g.\ Heavens \& Peacock 1988)
and these studies are generally in good agreement with numerical simulations
(Barnes \& Efstathiou 1987). Ryden (1988) and Quinn \& Binney (1992)
investigated the rate at which the direction of infalling angular momentum
should slew by studying the angular momenta of individual spherical shells
of protogalactic material. Quinn \& Binney found that the angular momenta
acquired by shells that differ in radius by a factor 2 have a clear tendency
to be antiparallel. Moreover, the angular momentum per unit mass of a shell
rises strongly with radius, with the consequence that the net angular
momentum of a galaxy tends to be aligned with the angular momentum of the
most recently accreted shell. These two results together imply that the net
spin axis of a halo tends to slew through more than $90\deg$ in the time
required for the radius of the currently accreting shell to double. This
time depends on the cosmology and the initial density profile (e.g. Fillmore
\& Goldreich 1984). For critical cosmic density, $\Omega=1$, it is
typically comparable to the current Hubble time and the halo's spin axis is
likely to slew by $\gta7\deg$ in the $0.9\Gyr$ that the Jiang \& Binney
simulation lasted. 

\section{Infalling satellites}

It has long been recognized that the Magellanic Clouds are spiralling into
the Milky Way. Dynamical models of this process (Murai \& Fujimoto 1980;
Gardiner et al.\ 1994; Moore \& Davis, 1994; Lin et al.\ 1995) have been
successful in predicting the proper motions of the Clouds, which have now
been measured to reasonable accuracy (Jones, Klemola \& Lin, 1994; Kroupa \&
Bastian, 1997).  These models are based on the assumption that the
Magellanic Stream comprises material that has been tidally stripped out of
the Clouds. Hence the success of these models implies that as the Clouds
sink deeper into the Galactic halo, a polar ring will form. Since there will
be more angular momentum in this ring than there is in the Galactic disk,
the orbit of the Clouds constitutes direct evidence for the accretion of
mis-aligned angular momentum.

Ibata, Gilmore \& Irwin (1994) discovered that the Galaxy has a much nearer satellite
than the Clouds, namely the Sagittarius Dwarf galaxy, that is almost hidden
from us by the Galactic centre even though it is at a Galactocentric
distance of only $16\kpc$. Like the Clouds, the Sgr Dwarf
is in a nearly polar orbit (Ibata et al.\ 1997), but the poles of the two orbits make an
angle of $\sim90\deg$ with one another. It seems that the Dwarf's orbit has
a remarkably short period $\sim1\Gyr$ (Ibata \& Lewis 1998). 

On such a short-period orbit the Dwarf is severely tidally limited by the
Galaxy, and there is direct observational evidence that the Dwarf is being
tidally shredded: an arc of associated material has now been detected that
extends over $\gta60\deg$ on the sky (Mateo, Olszewski \& Morrison, 1999).
Several authors have concluded that a Dwarf that contained only the observed
luminous material could not have survived tidal shredding for a Hubble time
on its present orbit. By enveloping the observed dwarf in a rather
homogeneous cloud of dark matter, Lewis \& Ibata (1998) were able to
construct a model of the Dwarf which retained 46\% of its original mass
after a Hubble time on its present orbit. 

Jiang \& Binney (this volume) ask how the Dwarf could have got into its
present configuration: galaxies are not likely to form in regions where an
external tidal field is strong, because the field would shear away
protogalactic material and prevent it accumulating locally. Moreover, the
configuration proposed by Lewis \& Ibata is finely tuned in the sense that
the dark halo has to be very homogeneous and sharp-edged. For these reasons
it seems likely that the Dwarf was formed at a considerable distance from
the Milky Way, and has subsequently moved onto its present tight orbit by
one of two mechanisms. Zhao (1998) suggested that the Dwarf was recently
scattered onto its orbit by an encounter with the Clouds. The problem with
this proposal is the softness of the potentials of the Dwarf and the Clouds
relative to the large velocity of any encounter between them: the required
scattering is through a substantial angle. Dynamical friction is the other
mechanism that could have moved the Dwarf onto a short-period orbit. The
problem here is that the present mass of the Dwarf, even including the Lewis
\& Ibata dark halo, is too small ($\sim10^9\msun$) for dynamical friction to
be effective.

Jiang \& Binney combine $N$-body simulations, in which both Dwarf and Galaxy
are live systems, with a semi-analytic model, that incorporates dynamical
friction and tidal limitation, to explore the rather large parameter space
of possible Dwarf histories. They identify a one-parameter family of initial
configurations in which the Dwarf starts out at ever greater Galactocentric
distances. In the most distant initial configuration explored, the Dwarf
starts from $R=250\kpc$ as an object of mass $10^{11}\msun$, largely in the
form of a dark halo with central velocity dispersion $\sim50\kms$ and
roughly constant circular speed.  At the other extreme, the
Dwarf starts from $R\simeq60\kpc$ with mass $\sim1.4\times10^{10}\msun$. The
off-axis angular momentum that the Dwarf brings to the Galaxy, which varies
by a factor $\sim30$ between extreme configurations, would have a pronounced
effect of the Galactic warp at the upper end of the range. Hence, it should
be possible to constrain the history of the Dwarf by modelling
the combined effect of the Dwarf and the Clouds on the outer disk.

\section{Conclusions}

From the kinematics of M31 we know that the Local Group has yet to
virialize. Hence, there is an a-priori expectation that out towards the
edges of the Local Group there is a reservoir of material from which M31 and
the  Galaxy are accreting at a significant rate.

There is a real possibility that this reservoir can be traced through
high-velocity clouds (HVCs). This proposition is controversial because the
HVCs form a heterogeneous group. Some are clearly associated with M31, the
Magellanic Clouds, and other, lower-mass systems such as Phoenix (St-Germain
et al, 1999). Others are material in the Galactic halo, but one can identify
substantial subset of HVCs that are most naturally interpreted as rather
massive objects of order $1\Mpc$ from the Galaxy that have yet to fall to
the centre of the Local Group. These clouds may together contain
$\sim10^{10}\msun$ of HI and enable the Galaxy to accrete HI at a rate of
$\sim0.8\msun\yr^{-1}$ and dark matter at a rate ten times higher.

Currently we know of no viable mechanism by which warps could survive long
in an isolated galaxy, and there is no longer observational support for the
proposition that warps exist in isolated galaxies. It seems likely that
warps are a manifestation of a galaxy accreting material whose net angular
momentum is not parallel to that of the galaxy. To generate a typical warp,
off-axis angular momentum must be accreted at such a rate that the spin axis
of the inner galaxy slewed by $\sim10\deg$ per Gyr.

Observations of the Magellanic Stream and the Sgr Dwarf Galaxy tell us that
the Milky Way is certainly accreting off-axis angular momentum. Whether this
accretion is fast enough to slew the inner Galaxy's spin axis by
$\sim10\deg$ per Gyr depends on how much dark-matter is being accreted
alongside the observed luminous matter. If there is 10 times as much dark as
luminous matter, it should be possible to explain the Galactic warp,
although the very peculiar morphology of the Galactic warp has yet to emerge
from a model of Galactic accretion.

\end{document}